# Three-dimensional imaging of buried heterointerfaces

Colum M. O'Leary[1,†], Haozhi Sha[1†], Jianhua Zhang[2,†], Cong Su[3,4,5], Salman Kahn[3,4,5], Huaidong Jiang[2], Alex Zettl[3,4,5], Jim Ciston[6] and Jianwei Miao[1*]

[1]*Department of Physics & Astronomy and California NanoSystems Institute, University of California, Los Angeles, CA 90095, USA.*

[2]*School of Physical Science and Technology, ShanghaiTech University, Shanghai 201210, China.*

[3]*Department of Physics, University of California at Berkeley, Berkeley, CA 94720, USA.*

[4]*Materials Sciences Division, Lawrence Berkeley National Laboratory, Berkeley, CA 94720, USA*

[5]*Kavli Energy NanoSciences Institute at the University of California, Berkeley, CA 94720, USA.*

[6]*National Center for Electron Microscopy Facility, Molecular Foundry, Lawrence Berkeley National Laboratory, Berkeley, CA 94720, USA.*

**We report three-dimensional (3D) structure determination of a twisted hexagonal boron nitride (h-BN) heterointerface from a single-view data set using multislice ptychography. We identify the buried heterointerface between two twisted h-BN flakes with a lateral resolution of 0.57 Å and a depth resolution of 2.5 nm. The latter is a significant improvement (~2.7 times) over the aperture-limited depth resolution of incoherent imaging modes such as annular-dark-field scanning transmission electron microscopy. This is attributed to the diffraction signal extending beyond the aperture edge with the depth resolution set by the curvature of the Ewald sphere. Future advances to this approach could improve the depth resolution to the sub-nanometer level and enable the identification of individual dopants, defects and color centers in twisted heterointerfaces and other materials.**

The development of twisted van der Waals (vdW) heterostructures, where individual layers of two-dimensional (2D) materials are successively stacked at relative rotation angles, has led to exciting applications in optoelectronics, sensing, energy storage, quantum computing and more [1,2]. The heterointerfaces in these materials are hosts of several exotic properties and phenomena such as unique band gaps [3,4], superconducting behavior [5,6] and confined photon emission [7,8]. While aberration-corrected electron microscopy can routinely image 2D materials at sub-angstrom resolution [9], it remains a challenge for high-resolution 3D imaging of buried heterointerfaces. Currently, the highest-resolution 3D imaging method is atomic electron tomography (AET) [10,11], which acquires a tilt series of projections from a sample by annular dark-field (ADF)-scanning transmission electron microscopy (STEM) and uses powerful algorithms to reconstruct the 3D structure of the sample. The 3D positions of individual atoms can then be localized with high precision [12]. AET has been advanced to determine the 3D atomic structure of crystal defects and amorphous materials [13-17]. However, current AET workflows rely on obtaining a number of ADF-STEM images at different sample orientation, which offers weak signal of light elements (i.e. B, C, N) abundant in 2D materials. While ptychographic AET (pAET) has recently been proposed to resolve the 3D atomic positions of low-Z elements [18], the experimental implementation of the method is still in its infancy [19]. The second 3D imaging approach is based on the principle of confocal microscopy, in which ADF-STEM is used to acquire a focal series from a single angle to reveal 3D structure with a depth resolution of several nanometers [20-23]. The depth resolution is limited by the convergence semi-angle subtended by the probe-forming aperture and is typically 1-2 orders of magnitude lower than the corresponding lateral resolutions [20-23]. The third approach is based on the principle of ankylography [24], a coherent diffractive imaging method [25], which requires neither tilting of the sample nor scanning of the focus position. Although it was controversial when ankylography was first published [26,27], this method of 3D structure determination from a single view has since inspired the development of multislice ptychography (MSP) [28-30] and other approaches using electrons, x-rays and visible light [31-38].

In this Letter, we advance MSP to determine the 3D structure of a buried heterointerface of two large, twisted h-BN crystals from a single-view data set. We achieve a lateral resolution of 0.57 Å and a depth resolution of 2.5 nm. The latter represents the highest depth resolution in any 3D imaging method from a single view, and is set by the maximum diffraction angle instead of the convergence semi-angle, which is in accordance with ankylography [24]. The experimental confirmation that the depth resolution is set by the maximum diffraction angle could significantly enhance the capability of MSP for high-resolution 3D imaging of heterointerfaces and other materials.

Figure 1 shows a schematic of the experimental setup for MSP on twisted h-BN crystals, in which a probe is scanned in a raster across the sample and 2D diffraction patterns are collected by a fast pixelated detector. The resulting data acquired from the scan comprises a single-view four-dimensional ptychographic array, which is used as input to an iterative phase retrieval algorithm [25,39,40]. Unlike typical STEM imaging techniques such as bright-field, annular bright/dark-field and differential phase contrast, the ptychographic reconstruction utilizes both the bright-field signal that extends to the semi-convergence angle ($\alpha$) as well as the dark-field signal up to the maximum collection angle ($\theta$) [41,42]. Furthermore, the reconstruction incorporates the inverse multi-slice approach for overcoming dynamical scattering from thick samples [28-30]. The output of the algorithm is the complex transmission function of the object for each slice of the specimen, as well as the complex function of the probe, which can be decomposed into several orthogonal modes to account for partial coherence effects [43–45]. The phase of the transmission function is highly sensitive to slight disturbances in the electron wave, and can be plotted to generate high-fidelity reconstructions of light-element species [46,47] with significantly higher light-element contrast than ADF-STEM imaging. In recent studies, MSP has been used to reveal structural inhomogeneities in single crystals [30,48,49] and distinguish individual nanotubes in projection [29,50,51]. Here, we demonstrate the application of MSP to stacked 2D materials. Upon MSP reconstruction, the top ($\psi_{top}$) and bottom ($\psi_{bot}$) crystals in the twisted system can be visualized independently from separate phase slices (Fig. S1 [52]).

To demonstrate the application of MSP to vdW heterointerfaces, we performed a ptychographic experiment on two twisted h-BN flakes [7] deposited onto a holey Si grid [Fig. 1(a)]. The experimental data was collected using the TEAM I double-corrected S/TEM instrument at the National Center for Electron Microscopy (voltage: 300kV, convergence semi-angle: 17.1 mrad, see Supplemental text [52]). A focused electron probe was scanned across the h-BN sample, and the transmitted diffraction pattern was collected for each position using a Gatan K3 fast pixelated detector (512×512 px, dwell time: 0.87 ms) operated in electron-counting mode. A total of 256×256 diffraction patterns were acquired with a step size of 0.25 Å between adjacent scan positions, resulting in a cumulative electron dose of $4.5 \times 10^6$ e$^-$ Å$^{-2}$. The data was collected over one of the holes in the sample grid to minimize background intensity, as highlighted with the dashed circle in Fig. 1(b). An ADF image acquired over the crystal structure, displayed in Fig. 1(c), clearly shows the atomic columns of both crystals. Furthermore, the averaged diffraction pattern displayed in Fig. 1(d) shows several diffraction disks interfering with the zeroth-order disk, indicating that both crystal flakes in the twisted h-BN are contributing to the diffraction intensity measurements. The relative in-plane twist angle $\theta_t$ between the crystals was determined as 19.05° from the Fourier transform of seven interface slices (Fig. S2 [52]). In recent studies, highly twisted h-BN interfaces ($\theta_t$ >14°) have demonstrated strong photon emission in the UV regime indicative of the presence of nitrogen color centers, thus the capability to decipher the location of highly twisted interfaces is of notable importance for quantum sensing and computing applications [7]. Direct correlation between sites of photon emission and the location of nitrogen vacancies is beyond the scope of this work and will be performed in future studies.

The ptychographic reconstructions were performed using the maximum-likelihood multislice engines available in the fold-slice package, which has been adapted from the Ptychoshelves package [30,53,54]. A 128×128 scan region was chosen from the original data set, and the detector was binned to 128×128 pixels before performing the reconstruction. Next, the complex object function was calculated using two separate reconstruction runs: (1) a single-slice ptychography reconstruction; and (2) an MSP reconstruction incorporating 34 slices, each separated by 0.5 nm. For each reconstruction, the complex probe was calculated for eight orthogonal modes (Fig. S3 [52]). A slice spacing of 0.5 nm was chosen in order to maximize depth resolution while minimizing the crosstalk of low spatial frequencies [55]. To further mitigate crosstalk, a regularization between slices was performed (Supplementary Text C) [52]. A 3D weighting function, $W$, was applied to the object in reciprocal space to mitigate artifacts caused by crosstalk of low spatial frequencies between adjacent reconstruction slices [30,55], $W = 1 - \frac{\tan^{-1}\left(\left[\gamma \frac{k_z}{k_r + \epsilon}\right]^2\right)}{\pi/2}$, where $k_r$ and $k_z$ correspond to the spatial frequencies in the lateral and depth directions, $\epsilon$ is a small number to prevent division-by-zero errors, and γ is a regularization parameter. A visualization of $W$ is shown for the 2D case ($k_x$-$k_z$) in Fig. S4 [52]. By increasing γ, information transfer at $k_z > k_r$ is dampened, resulting in an object reconstruction with fewer artifacts at the expense of lower depth resolution. In this work, γ was set to 0.5 (Table S1 [52]). Applying values of γ below 0.5 resulted in increased artifacts in the reconstruction. A full list of parameters for the reconstructions performed are listed in Table S1, and several parameters are described in the Supplementary Note 3 [52].

To demonstrate the improvement of lateral resolution using MSP, the reconstructed phase of the object functions for the single-slice and multislice reconstructions are shown in Figs. 2(e) and (g), respectively. The thickness of the stacked h-BN crystals was estimated as 12 nm (Fig. S5 [52]), and the summed-multislice result here was obtained from 24 slices of the object reconstruction. It is evident that the width of the atomic column potentials is significantly reduced for the case of MSP. To compare both reconstructions, the same region of interest in Figs. 2(e) and (g), enclosed by the dashed square, is magnified inset. The MSP reconstruction reveals several atomic column potentials that could not be observed using the single-slice approach, which suggests that the dynamical scattering effects have been accounted for using the multislice approach. For further comparison, the Fourier transforms of the single- and multislice approaches are shown in Figs. 2(f) and (h), respectively. It is evident that the multislice approach resolves diffraction spots out to 0.57 Å, while the single-slice approach can resolve spots out to 0.72 Å. In addition, there is a noticeable improvement of the probe reconstruction using MSP (Fig. S3 [52]). For the single-slice approach, the higher order modes are corrupted with artifacts, possibly due to the dynamical scattering. In the case of the multislice reconstruction, the probe intensity is centered about the origin, and the fringes are clearly resolved in the phase of all eight modes. This improvement of MSP over single-slice ptychography is significant because, while h-BN columns would not be expected to cause significant dynamical scattering at such crystal thicknesses with 300kV electrons, there are noticeable improvements to the reconstruction if such scattering is accounted for.

Supplemental Video S1 and Figure S5 [52] show the propagation through all 34 reconstructed slices by MSP, demonstrating the depth-dependent structural variation of the twisted h-BN crystals. To find the location of the buried interface, we display five 1-nm-thick slices with the depth position ranging from 4 to 8 nm below the surface of the top flake. Figure 3 shows the phase images and the corresponding Fourier transforms of the five slices. The phase image of the 1$^{st}$ slice clearly shows the h-BN atomic structure of the top flake [Fig. 3(a)], while the weak Bragg peaks of the bottom flake are also visible [Fig. 3(f)]. Upon propagation towards the interface [Fig. 3(b) and (g)], the contrast of the real-space phase and the intensity of the Bragg peak of the bottom flake increase. At the depth position of 6 nm (i.e. the 3$^{rd}$ slice), atomic columns from both flakes are clearly visible from the phase slice in Fig. 3(c), and there are similar Bragg peak contributions as shown in Fig. 3(h). Upon inspection of all 34 slices (Fig. S5) the interface is estimated to be located 6 nm below the surface of the top h-BN flake. After propagation through the interface to the depth position of 7 nm [Fig. 3(d) and (i)], the relative contribution of the bottom flake increases, while the final displayed slice only clearly shows the bottom flake in the phase image [Fig. 3(e)], with weak Bragg peaks from the top flake [Fig. 3(j)]. Additionally, the x-z profiles of the multislice reconstruction are included in Supplemental Video S2 (see Supplemental text [52]).

Next, we estimate the depth resolution of the MSP reconstruction based on a knife-edge analysis. The coordinates of the atomic columns for the bottom flake are determined using the peak fitting functions of the Atomap python library [56]. In order to determine the area of each column peak, a sum of the reconstructed slices of the top flake (Fig. S1 [52]) is used as a basis for masking. The mask is applied to each phase slice and, for each atomic coordinate, the phase within the surrounding mask is summed and registered as a function of thickness. The average phase thickness profile is determined from 100 individual atomic columns, and the normalized result of a representative column is plotted in Fig. 4(a). The reconstructed phase slices I-IV shown in Fig. 4(b) highlight the propagation of the peak phase intensity from the vacuum above the sample into the top flake. Using linear interpolation of the scatter plot, the knife-edge resolution is obtained by calculating the difference between the thickness values at 10% and 90% maximum phase. The average depth resolution based on the 100 columns is determined to be 2.5 nm. This is consistent with previous numerical simulation results [30] and is 2.7 times higher than the aperture-limited depth resolution of 6.7 nm, calculated by $d_z \approx \frac{\lambda}{\alpha^2}$, where $d_z$ is the depth resolution, $\lambda$ is the wavelength and $\alpha$ is the convergence semi-angle (Fig. 1) [52]. We attribute the significant resolution improvement to the diffraction signal beyond the bright-field disk. According to ankylography [24], the depth resolution is determined by,

$$d_z \approx \frac{2\lambda}{\theta^2} \qquad (1)$$

where $\theta$ is the maximum diffraction angle (Fig. 1) and the factor of 2 is because it is derived from coherent diffraction with a parallel incident beam assumption. In our experimental data, the maximum diffraction angle is 36.2 mrad [Fig. 1(d)], which corresponds to a depth resolution of 3 nm and is more consistent with our depth-resolution measurement of 2.5 nm. The slightly better measured depth resolution is due to the parallel incident beam assumption in Eq. (1). If we consider the convergence semi-angle in the experiment, the maximum diffraction angle of some signals is actually larger than 36.2 mrad. To observe the improvement of depth resolution with maximum diffraction angle, the average knife-edge resolution was plotted for 4 separate reconstructions with different angular cutoff values [Fig. 4(c)]. The calculated resolution and standard deviation (error bars) decrease for larger cutoff angles up to the maximum diffraction angle of 36.2 mrad. Upon padding the diffraction patterns to 45.2 mrad before reconstruction, the variance of the resolution further decreases, but the mean resolution does not improve. It should be noted that the improvement of the depth resolution requires the increase of the probe overlap and/or the oversampling of the diffraction patterns so that the number of independently measured points is larger than the that of the unknown variables [57]. In this experiment, the overdetermination ratio ($\sigma$) between the measured points and the unknown variables is 7.38 (Supplemental text [52]), and it is expected that larger $\sigma$ will result in the diffraction-dependent resolution improvement [Fig. 4(c)] converging to that expected from Eq. (1). Although MSP can overcome thickness limitations imposed by most atomic-resolution S/TEM techniques, there are several limitations to this approach. Firstly, while MSP can provide depths of field beyond those of incoherent imaging techniques such as ADF-STEM, there is a limit to the maximum thickness (i.e. <100 nm) that can be used, as a result of significant electron absorption and dynamical scattering [30,49]. Nevertheless, we anticipate that the workflow shown here can be used to identify interfaces in light-element vdW heterostructures with cumulative thicknesses up to 50 nm. Secondly, as the number of slices increases, $\sigma$ decreases and mixing of low spatial frequencies between adjacent slices is significant, eventually leading to an ill-posed problem with many unknown variables. This is compounded by the coincidence loss suffered by large-array pixelated detectors when high electron currents are used. Thus, the use of a strong regularization to symmetrize adjacent slices is crucial for increasing the fidelity of each phase reconstruction slice and improving convergence of the algorithm. However, once the regularization is weakened for improved depth resolution, the crosstalk of low spatial frequencies may give rise to spurious intensity values, thus limiting the phase sensitivity of this approach for single-atom identification. Potential techniques to overcome this issue include using pixelated detectors with larger pixel size; acquiring several ptychographic data sets from a range of geometric aberration values to aid convergence; and performing data acquisition over a range of projection angles.

In summary, we have demonstrated the use of multislice ptychography for probing interfaces of twisted crystals on the nanoscale from a single view. The depth-resolution improvement over the aperture limit demonstrates the potential to probe crystals in projection at the sub-nanometer level. With further depth-resolution improvement, either via increased convergence and diffraction angles using state-of-the-art aberration correctors [58] or through tomographic methods [19,59], structural inhomogeneities at the sub-nanometer level should be achievable, providing important insights into the crystal order/disorder of complex 2D materials.

This research was primarily supported by the US Department of Energy, Office of Science, Basic Energy Sciences, Division of Materials Sciences and Engineering under award no. DE-SC0010378. C.M.O. and H.S. acknowledge support by STROBE: a National Science Foundation Science and Technology Center under award no. DMR-1548924 and the Army Research Office Multidisciplinary University Research Initiative (MURI) program under grant no. W911NF-18-1-0431. The electron microscopy experiments were performed at the Molecular

Foundry, which is supported by the Office of Science, Office of Basic Energy Sciences of the US Department of Energy under contract no. DE-AC02-05CH11231. C.S. and A.Z. acknowledge the support by the Director, Office of Science, Office of Basic Energy Sciences, Materials Sciences and Engineering Division, of the US Department of Energy under contract no. DEAC02-05-CH11231, within the sp2 -Bonded Materials Program (KC-2207) which provided for preliminary TEM and Raman characterization of the h-BN material, and by the van der Waals Heterostructures program (KCWF16) which provided for assembly of the twisted h-BN material. J. Z. and H. J. acknowledge the support by the Major State Basic Research Development Program of China ( 2022YFA1603703). Data processing was carried out using the cSAXS ptychography MATLAB package developed by the Science IT and the coherent X-ray scattering (CXS) groups, Paul Scherrer Institut, Switzerland.

[†]These authors contributed equally to this work.

[*]miao@physics.ucla.edu

**Figures and Figure Captions**

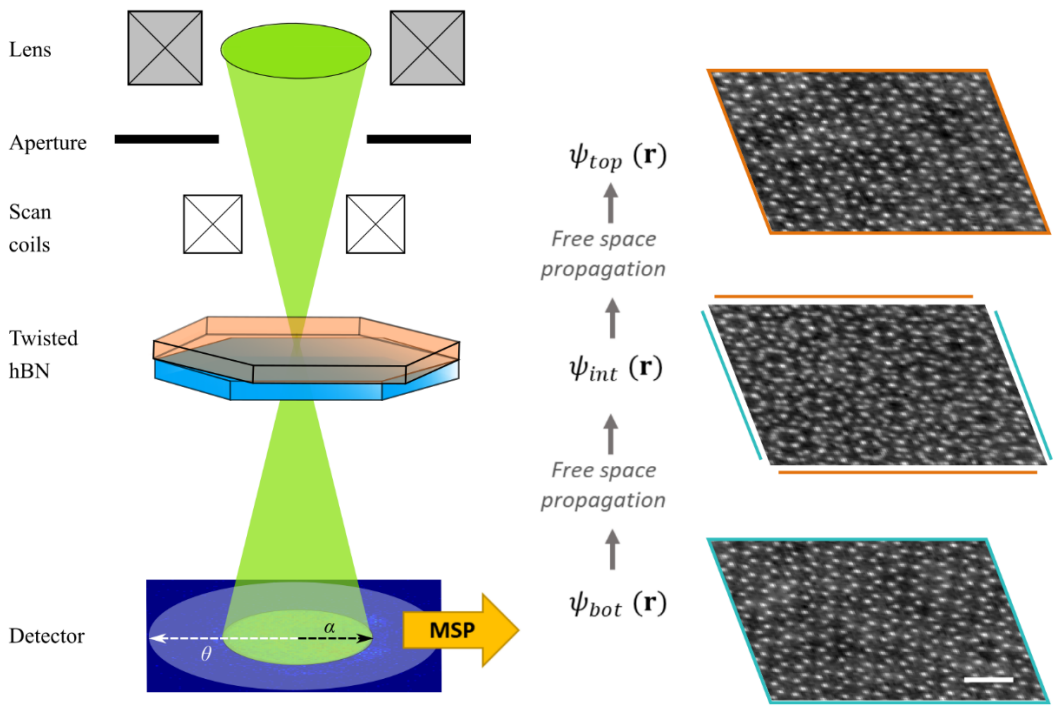

**Figure 1.** Schematic of multi-slice ptychography (MSP) on twisted h-BN crystals. At each point in a raster scan across the sample, a diffraction pattern is collected by a fast pixelated detector. The bright-field signal collected by the detector is highlighted in green with convergence semi-angle α, while the dark-field signal comprises the annulus shaded in pale blue with maximum diffraction angle θ. Reconstruction via MSP provides atomic-scale phase reconstructions of different sample slices, revealing nanoscale depth information. Here, each phase slice shown is generated from the sum of seven separate 0.5-nm slices (Fig. S1 [52]). Scale bar, 0.5 nm.

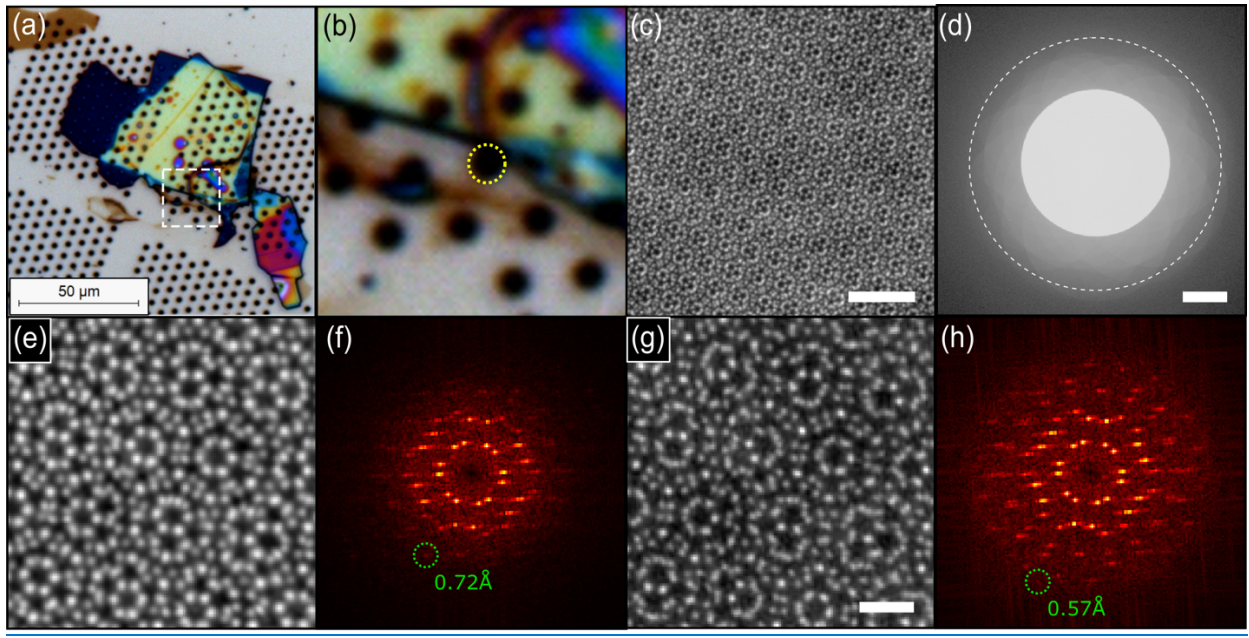

**Figure 2. Ptychographic reconstruction of twisted h-BN crystals.** (a) Optical microscope image of twisted h-BN flakes. (b) A magnified view of the region enclosed by the white square in (a) shows two twisted flakes. The dashed circle indicates the grid hole used for data collection. (c) Annular dark-field (ADF) image of the twisted h-BN flakes. (d) Averaged diffraction pattern (logarithmic scale) from the ptychographic data set, where the maximum diffraction angle is 36.2 mrad. Scale bar, 10 mrad. (e) Ptychographic phase obtained from single-slice ptychography. (f) Magnitude of the Fourier transform of (e), showing a lateral resolution of 0.72 Å. (g) Ptychographic phase obtained from the sum of 24 0.5-nm-thick slices reconstructed via multislice ptychography. (h) Magnitude of the Fourier transform of (g), showing a lateral resolution of 0.57Å. Scale bars, 2 nm (c) and 0.5 nm (g).

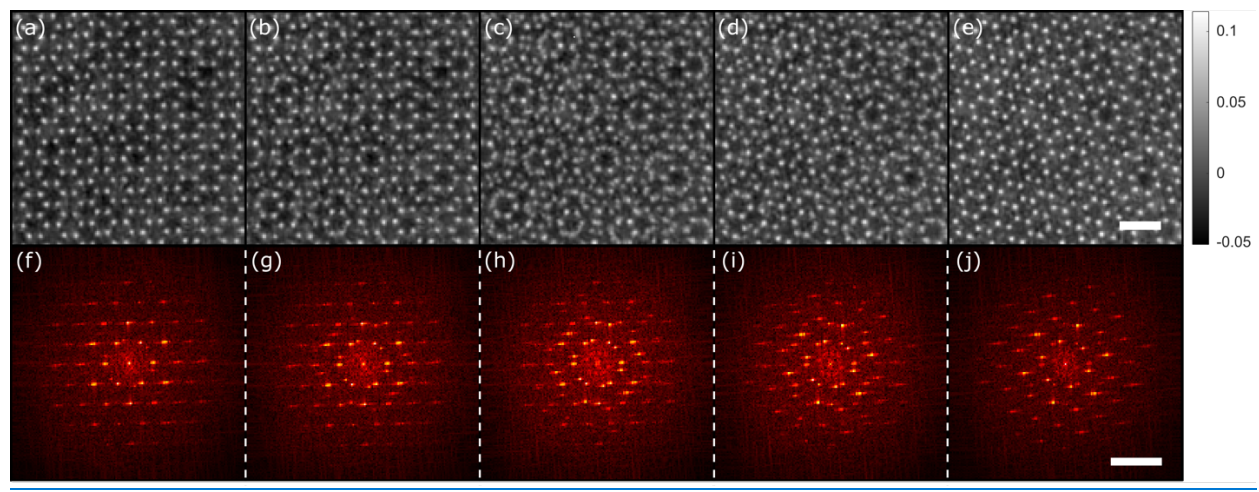

**Figure 3. Probing a buried interface in the twisted h-BN flakes.** Reconstructed phase slices of the twisted h-BN flakes at depth positions of (a) 4nm, (b) 5nm, (c) 6nm, (d) 7nm, and (e) 8nm below the surface of the top flake. Scale bar, 0.5 nm. (f)-(j) The Fourier transforms of the corresponding five phase images. Scale bar, 10 nm$^{-1}$. Color scale, phase (radians).

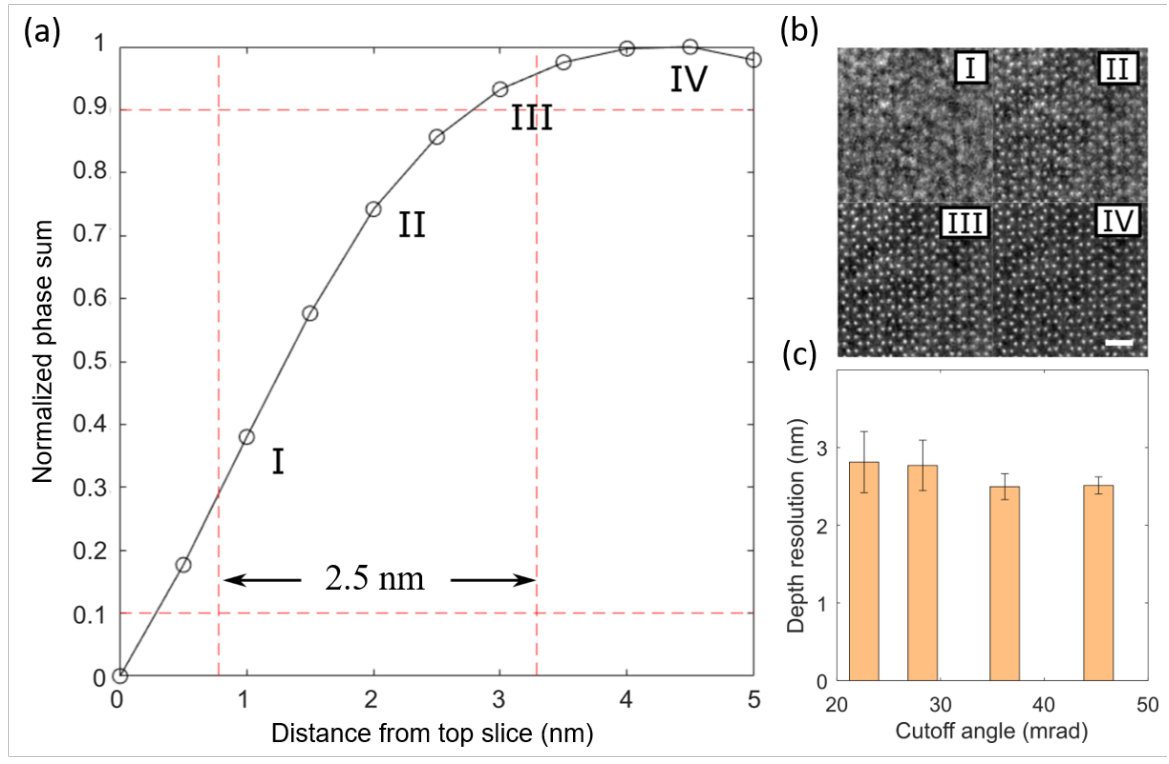

**Figure 4. Knife-edge plot for the twisted h-BN crystals.** Representative knife edge for an atomic column in the top h-BN flake. A linear interpolation is applied to the measured values to calculate the knife-edge resolution corresponding to the z-distance between 10% and 90% maximum phase. (b) Reconstructed phase slices at the corresponding thickness values labelled on the knife-edge plot. Scale bar: 0.5 nm. (c) Average depth resolution as a function of angular cutoff used in the reconstruction. The measured depth resolution is $2.81\pm0.79$ nm, $2.77\pm0.65$ nm, $2.5\pm0.33$ nm and $2.51\pm0.22$ nm for a maximum cutoff angle of 22.6, 28.3, 36.2 and 45.2 mrad, respectively. In the case when the diffraction patterns were padded with zeros to 45.2 mrad before reconstruction, the variance of the resolution decreases, but the mean resolution does not improve.